\documentclass[aps,prc,twocolumn,superscriptaddress, showkeys,nofootinbib]{revtex4-1}  

\usepackage[utf8]{inputenc}
\usepackage{graphicx,color}
\usepackage{epstopdf}
\usepackage{hyperref}

\usepackage{xspace}	
\newcommand{\roots} {\mbox{$\sqrt{s_{\rm NN}}$}\xspace}
\newcommand{\GeVc} {\mbox{GeV/$\textit{c}$}\xspace}
\def  \vn          {\mbox{$\textit{v}_{n}$  }\xspace}

\begin{document}
\title{Collision-System and Beam-Energy Dependence of Anisotropic Flow Fluctuations}
\affiliation{Abilene Christian University, Abilene, Texas   79699}
\affiliation{AGH University of Science and Technology, FPACS, Cracow 30-059, Poland}
\affiliation{Alikhanov Institute for Theoretical and Experimental Physics NRC "Kurchatov Institute", Moscow 117218}
\affiliation{Argonne National Laboratory, Argonne, Illinois 60439}
\affiliation{American University of Cairo, New Cairo 11835, New Cairo, Egypt}
\affiliation{Brookhaven National Laboratory, Upton, New York 11973}
\affiliation{University of California, Berkeley, California 94720}
\affiliation{University of California, Davis, California 95616}
\affiliation{University of California, Los Angeles, California 90095}
\affiliation{University of California, Riverside, California 92521}
\affiliation{Central China Normal University, Wuhan, Hubei 430079 }
\affiliation{University of Illinois at Chicago, Chicago, Illinois 60607}
\affiliation{Creighton University, Omaha, Nebraska 68178}
\affiliation{Czech Technical University in Prague, FNSPE, Prague 115 19, Czech Republic}
\affiliation{Technische Universit\"at Darmstadt, Darmstadt 64289, Germany}
\affiliation{ELTE E\"otv\"os Lor\'and University, Budapest, Hungary H-1117}
\affiliation{Frankfurt Institute for Advanced Studies FIAS, Frankfurt 60438, Germany}
\affiliation{Fudan University, Shanghai, 200433 }
\affiliation{University of Heidelberg, Heidelberg 69120, Germany }
\affiliation{University of Houston, Houston, Texas 77204}
\affiliation{Huzhou University, Huzhou, Zhejiang  313000}
\affiliation{Indian Institute of Science Education and Research (IISER), Berhampur 760010 , India}
\affiliation{Indian Institute of Science Education and Research (IISER) Tirupati, Tirupati 517507, India}
\affiliation{Indian Institute Technology, Patna, Bihar 801106, India}
\affiliation{Indiana University, Bloomington, Indiana 47408}
\affiliation{Institute of Modern Physics, Chinese Academy of Sciences, Lanzhou, Gansu 730000 }
\affiliation{University of Jammu, Jammu 180001, India}
\affiliation{Joint Institute for Nuclear Research, Dubna 141 980}
\affiliation{Kent State University, Kent, Ohio 44242}
\affiliation{University of Kentucky, Lexington, Kentucky 40506-0055}
\affiliation{Lawrence Berkeley National Laboratory, Berkeley, California 94720}
\affiliation{Lehigh University, Bethlehem, Pennsylvania 18015}
\affiliation{Max-Planck-Institut f\"ur Physik, Munich 80805, Germany}
\affiliation{Michigan State University, East Lansing, Michigan 48824}
\affiliation{National Research Nuclear University MEPhI, Moscow 115409}
\affiliation{National Institute of Science Education and Research, HBNI, Jatni 752050, India}
\affiliation{National Cheng Kung University, Tainan 70101 }
\affiliation{Nuclear Physics Institute of the CAS, Rez 250 68, Czech Republic}
\affiliation{Ohio State University, Columbus, Ohio 43210}
\affiliation{Institute of Nuclear Physics PAN, Cracow 31-342, Poland}
\affiliation{Panjab University, Chandigarh 160014, India}
\affiliation{Pennsylvania State University, University Park, Pennsylvania 16802}
\affiliation{NRC "Kurchatov Institute", Institute of High Energy Physics, Protvino 142281}
\affiliation{Purdue University, West Lafayette, Indiana 47907}
\affiliation{Rice University, Houston, Texas 77251}
\affiliation{Rutgers University, Piscataway, New Jersey 08854}
\affiliation{Universidade de S\~ao Paulo, S\~ao Paulo, Brazil 05314-970}
\affiliation{University of Science and Technology of China, Hefei, Anhui 230026}
\affiliation{Shandong University, Qingdao, Shandong 266237}
\affiliation{Shanghai Institute of Applied Physics, Chinese Academy of Sciences, Shanghai 201800}
\affiliation{Southern Connecticut State University, New Haven, Connecticut 06515}
\affiliation{State University of New York, Stony Brook, New York 11794}
\affiliation{Instituto de Alta Investigaci\'on, Universidad de Tarapac\'a, Arica 1000000, Chile}
\affiliation{Temple University, Philadelphia, Pennsylvania 19122}
\affiliation{Texas A\&M University, College Station, Texas 77843}
\affiliation{University of Texas, Austin, Texas 78712}
\affiliation{Tsinghua University, Beijing 100084}
\affiliation{University of Tsukuba, Tsukuba, Ibaraki 305-8571, Japan}
\affiliation{Valparaiso University, Valparaiso, Indiana 46383}
\affiliation{Variable Energy Cyclotron Centre, Kolkata 700064, India}
\affiliation{Warsaw University of Technology, Warsaw 00-661, Poland}
\affiliation{Wayne State University, Detroit, Michigan 48201}
\affiliation{Yale University, New Haven, Connecticut 06520}

\author{M.~S.~Abdallah}\affiliation{American University of Cairo, New Cairo 11835, New Cairo, Egypt}
\author{J.~Adam}\affiliation{Brookhaven National Laboratory, Upton, New York 11973}
\author{L.~Adamczyk}\affiliation{AGH University of Science and Technology, FPACS, Cracow 30-059, Poland}
\author{J.~R.~Adams}\affiliation{Ohio State University, Columbus, Ohio 43210}
\author{J.~K.~Adkins}\affiliation{University of Kentucky, Lexington, Kentucky 40506-0055}
\author{G.~Agakishiev}\affiliation{Joint Institute for Nuclear Research, Dubna 141 980}
\author{I.~Aggarwal}\affiliation{Panjab University, Chandigarh 160014, India}
\author{M.~M.~Aggarwal}\affiliation{Panjab University, Chandigarh 160014, India}
\author{Z.~Ahammed}\affiliation{Variable Energy Cyclotron Centre, Kolkata 700064, India}
\author{I.~Alekseev}\affiliation{Alikhanov Institute for Theoretical and Experimental Physics NRC "Kurchatov Institute", Moscow 117218}\affiliation{National Research Nuclear University MEPhI, Moscow 115409}
% ...

\collaboration{STAR Collaboration}\noaffiliation
%
\begin{abstract}
Elliptic flow measurements from two-, four- and six-particle correlations are used to investigate 
flow fluctuations in collisions of U+U at \roots = 193~GeV, Cu+Au at \roots = 200~GeV 
and Au+Au spanning the range \roots = 11.5 - 200~GeV. The measurements show a strong dependence of the flow fluctuations on collision centrality, a modest dependence on system size, and very little if any, dependence on particle species and beam energy. The results, when compared to similar LHC measurements, viscous hydrodynamic calculations, and T$\mathrel{\protect\raisebox{-2.1pt}{R}}$ENTo model eccentricities, indicate that initial-state-driven fluctuations predominate the flow fluctuations generated in the collisions studied.
\end{abstract}

\maketitle


A wealth of studies of heavy-ion collisions at the Relativistic Heavy Ion Collider (RHIC) and the Large Hadron Collider (LHC) indicate that an exotic state of matter,  called the Quark-Gluon Plasma (QGP), is created in the hot and dense environment present in these collisions. Ongoing studies at RHIC and the LHC are focused on developing a complete understanding of the dynamical evolution and the transport properties of the QGP.

Several analysis techniques have been employed to study the QGP. In particular, azimuthal anisotropy measurements
of the produced particles have been used to study the viscous hydrodynamic response of the QGP to the 
spatial distribution of the initial energy density  produced in the early stages of 
the collisions~\cite{Hirano:2005xf,Huovinen:2001cy,Hirano:2002ds,Romatschke:2007mq,Luzum:2011mm,Song:2010mg,Qian:2016fpi,Magdy:2017kji,Schenke:2011tv,Teaney:2012ke,Gardim:2012yp,Lacey:2013eia}.  
The azimuthal anisotropy of the particles produced relative to the flow planes $\Psi_{\rm n}$,  can be quantified  via Fourier  
decomposition~\cite{Voloshin:1994mz,Poskanzer:1998yz} of the distribution of their azimuthal angle ($\phi$):
\begin{eqnarray}
\label{eq:1-1}
\frac{dN}{d\phi}  &\propto&   1+ 2 \sum^{\infty}_{n=1}\vn \cos\left[  n (\phi - \Psi_{\rm n})   \right]   ,
\end{eqnarray}
where the first Fourier harmonic, $v_{1}$, is termed directed flow; $v_{2}$ is termed elliptic flow; $v_{3}$ is 
termed triangular flow, etc. 
A wealth of information on the characteristics of the QGP has been gained via 
 studies of directed and elliptic flow~\cite{PHOBOS:2006dbo,PHOBOS:2005ylx,STAR:2002hbo,STAR:2003xyj,STAR:2000ekf,PHENIX:2018wex,PHENIX:2002hqx,Adam:2019woz,Magdy:2018itt,ALICE:2011ab,ALICE:2012vgf,ALICE:2016ccg,ALICE:2016tlx,ATLAS:2018ezv,CMS:2012tqw,CMS:2012zex,CMS:2013jlh,CMS:2013wjq,ALICE:2010suc}, 
higher-order flow harmonics, $v_{n > 2}$~\cite{Alver:2010gr,PHENIX:2011yyh,PHENIX:2014uik,ATLAS:2015qwl,ATLAS:2017hap,Adare:2011tg,ALICE:2011ab,Adamczyk:2017ird,Magdy:2017kji,Adamczyk:2017hdl,Chatrchyan:2013kba,CMS:2019cyz}, flow fluctuations~\cite{PHOBOS:2010ekr,PHOBOS:2007wlf,Alver:2008zza,Ollitrault:2009ie,Alver:2010rt,Qiu:2011iv,ATLAS:2014qxy,CMS:2017glf,ALICE:2018rtz,PHENIX:2018lfu,ATLAS:2019peb}, the correlations between different flow harmonics~\cite{Aad:2015lwa,Adamczyk:2017hdl,STAR:2018fpo,Adam:2020ymj,ALICE:2021klf,ALICE:2021adw,ATLAS:2018ngv}, { and correlations of symmetry planes~\cite{Adam:2020ymj,Acharya:2017zfg,CMS:2019nct,Magdy:2022ize,Magdy:2022jai,CMS:2019nct,Yan:2015jma,Magdy:2021sba,ATLAS:2017rij}}.

Anisotropic flow driven by the spatial anisotropy of the initial-state energy density is characterized by the eccentricity vectors~\cite{Alver:2010dn,Petersen:2010cw,Lacey:2010hw,Teaney:2010vd,Qiu:2011iv}:
\begin{eqnarray}
\mathcal{E}_{n}  \equiv \varepsilon_{n} e^{i {{n}} \Phi_{n} } \equiv  
  - \frac{\int d^2r_\perp\, {r}^{n}\,e^{i {{n}} \varphi}\, \rho_{e}(r,\varphi)}
           {\int d^2r_\perp\, {r}^{n}\,\rho_{e}(r,\varphi)}, ~({n} ~>~ 1),
\label{epsdef1}
\end{eqnarray}
where  $\varepsilon_{n} = {\left< \left| \mathcal{E}_{n} \right|^2 \right>}^{1/2}$ and 
${\Phi_{n}}$ are the magnitudes and azimuthal directions of the eccentricity vectors, $\varphi$ is the spatial azimuthal angle, and $\rho_{\textit{e}}(r,\varphi)$ represents the initial anisotropic energy density profile~\cite{Teaney:2010vd,Bhalerao:2014xra,Yan:2015jma}. 

The $v_{2}$ and $v_{3}$ harmonics are, to a reasonable approximation, linearly related to the initial-state anisotropies, 
$\varepsilon_{{{2}}}$ and $\varepsilon_{{{3}}}$, respectively~\cite{Song:2010mg, Niemi:2012aj,Gardim:2014tya, Fu:2015wba,Holopainen:2010gz,Qin:2010pf,Qiu:2011iv,Gale:2012rq,Liu:2018hjh}: 
\begin{eqnarray}\label{eq:1-2}
v_{n} = \kappa_{n} \varepsilon_{n}, \, n=2,3,
\end{eqnarray}
where $\kappa_{n}$ encodes the medium response which is sensitive to the 
specific viscosity, {\it i.e.}, the ratio of dynamic viscosity to entropy density, $\eta/s$.
Precision extractions of $\eta/s$ require reliable model constraints for 
initial-state eccentricities and their fluctuations across a broad range of beam energies 
and collision systems \cite{Schenke:2019ruo,Alba:2017hhe}. Such constraints can be achieved via measurements 
of the flow harmonics and the event-by-event flow fluctuations for different systems and 
collision energies~\cite{Qiu:2011iv}. 

Flow fluctuations could arise from several underlying sources. They could develop in the initial state
 due to density fluctuations,   during hydrodynamic evolution due to dissipation, and during hadronization. 
The precise role of the initial-state eccentricity fluctuations has attracted 
considerable recent attention~\cite{Manly:2005zy,Magdy:2020gxf,Rao:2019vgy}. However,
the importance of the respective fluctuation sources has not been fully charted.

The multiparticle flow harmonics $v_n\{k\}$, {with cumulants order ${\it{k}}$=2, 4, and 6}, obtained via multiparticle correlation methods~\cite{Bilandzic:2013kga,Jia:2017hbm} can give direct access to the event-by-event flow fluctuations~\cite{Borghini:2000sa,Aaboud:2019sma}.
Consequently, extensive measurements of $v_n\{k\}$ for different collision systems and beam energies
could help to disentangle the fluctuation contributions from their respective sources, as well as 
establish whether flow fluctuations depend on the temperature, $T$, baryon chemical 
potential, $\mu_{\rm B}$, or both. 
It could also provide unique supplemental constraints to distinguish between 
different initial-state models and  reduce the fluctuations-related uncertainties associated 
with the extraction of $\eta/s(T, \mu_{\rm B})$.

In this letter, we report new flow fluctuation measurements in collisions of  U+U at \roots = 193~GeV, Cu+Au at \roots = 200~GeV and Au+Au spanning the range \roots = 11.5 - 200~GeV. The measurements are derived from the flow harmonics $v_2\{k\}$, extracted via multi-particle cumulants for ${\it{k}}$=2, 4, and 6. The extractions are comprehensive and benefit from consistent analysis across all systems and beam energies. Several of the extracted values for $v_2\{2\}$ and $v_2\{4\}$, used in the fluctuations measurements, are in good agreement with earlier charged hadron  $v_2\{2\}$ and $v_2\{4\}$ measurements for U+U (\roots = 193~GeV \cite{Adamczyk:2015obl}) and Au+Au collisions { spanning the range \roots = 11.5 - 200~GeV~\cite{Adamczyk:2012ku,Adams:2004bi}.}


The data reported in this analysis were recorded with a minimum-bias trigger using the STAR detector~\cite{Harris:1993ck}, with the low-energy Au+Au data being collected as a part of the STAR Beam Energy Scan (BES-I) program.
 The collision vertices were reconstructed using tracks measured with charged-particle trajectories detected in the STAR Time Projection Chamber (TPC) in a 0.5 T magnetic field pointing along the beam direction (z-axis)~\cite{Anderson:2003ur}.
%
%
%
Events were selected to be within a radius $r<2$~cm relative to the beam axis and within specific ranges of the center of the TPC in the direction along the beam axis, $v_{z}$ with the values $\pm$ 30~cm for U+U at \roots = 193~GeV, Cu+Au at \roots = 200~GeV, and Au+Au at \roots = 200~GeV, $\pm$ 40 cm at \roots = 54.4, 39, 27, 19.6~GeV and $\pm$ 50 cm at \roots = 11.5~GeV.

The collision centrality was determined via a Monte Carlo Glauber calculation tuned 
to match the event-by-event multiplicity measurements~\cite{Adamczyk:2012ku,Abelev:2009bw}.   
Analyzed tracks were required to have a Distance of Closest Approach (DCA) to the primary vertex of 
$<3$~cm, and to have more than 15 out of a possible 45 TPC space points used in their reconstruction. 
Furthermore, the ratio of the number  of fit-points used to the maximum possible number of TPC space points was required to be larger than 0.52 to remove split tracks.  The transverse momentum ($p_{\rm T}$) of the tracks was limited to $0.2 < p_{\rm T} < 4.0$~$\rm{GeV}$/$c$ for charged particles and to $0.2 < p_{\rm T} < 2.0$~$\rm{GeV}$/$c$ for the identified particle species. Particle identification (for pi, K, p) is based on the compound use of the ionization energy loss, dE/dx, in the TPC~\cite{STAR:1997sav}, and the squared mass from the TOF~\cite{Llope:2003ti} detector.

The operational framework of the multiparticle cumulant technique is given
in Refs.~\cite{Bilandzic:2010jr,Bilandzic:2013kga} and its extension to the method of subevent cumulants is summarized in Ref.~\cite{Jia:2017hbm}. 
Particle pairs, quadruplets and sextuplets were selected in the range $\mathrm{|\eta| < 1}$.
The $2m$-particle azimuthal correlator is obtained by averaging over all unique combinations in one event, then over 
all events~\cite{Borghini:2001vi}:
\begin{eqnarray}
\label{eq:c1}
\langle\langle 2m \rangle\rangle &=& \langle\langle e^{in\sum_{j=1}^{m}(\phi_{2j-1}-\phi_{2j})} \rangle\rangle,
\end{eqnarray}
to give the four- and six- particle cumulants as:
\begin{eqnarray}
\label{eq:c2-1}
c_n\{4\} &=& \left\langle\left\langle 4\right\rangle\right\rangle-2\left\langle\left\langle 2\right\rangle\right\rangle^2,\\
c_n\{6\} &=& \left\langle\left\langle 6\right\rangle\right\rangle-9\left\langle\left\langle 4\right\rangle\right\rangle\left\langle\left\langle 2\right\rangle\right\rangle+12\left\langle\left\langle 2\right\rangle\right\rangle^3.
\label{eq:c2-2}
\end{eqnarray}
%
%
 \begin{figure}[t]
 \vskip -0.6cm
 \includegraphics[width=0.9\linewidth, angle=0]{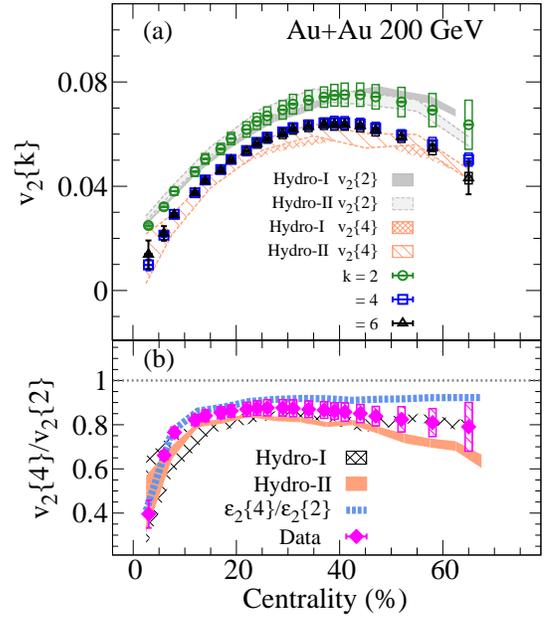}
\vskip -0.5cm
\caption{Comparison of the charged hadrons  two-, four- and six-particle elliptic flow, 
${v_{2}\lbrace k \rbrace}$, ${\it{k}}$ = 2, 4, and 6, panel (a), and the ratio,  
${v_{2}\lbrace 4 \rbrace / v_{2}\lbrace 2 \rbrace}$, panel (b),  vs. centrality,  
in the $p_{\rm T}$ range $0.2 - 4.0$~\GeVc for Au+Au collisions at \roots = 200~GeV;  
the range 1-5\% is used instead of 0-5\% \cite{STAR:2011ert} (see text).
The vertical lines and the open boxes indicate the respective statistical and systematic uncertainties.
The hatched bands and dashed curves represent the model calculations presented in 
Refs.~\cite{Schenke:2019ruo} (Hydro-I) and \cite{Alba:2017hhe} (Hydro-II), and the eccentricity ratio  
${\varepsilon_{2}\lbrace 4 \rbrace / \varepsilon_{2}\lbrace 2 \rbrace}$~\cite{Alba:2017hhe}, as indicated.
 }
 \label{Fig:1}
 \end{figure}
%
%
 \begin{figure*}[t]
  \vskip -0.5cm
\includegraphics[width=1.02\linewidth,angle=0]{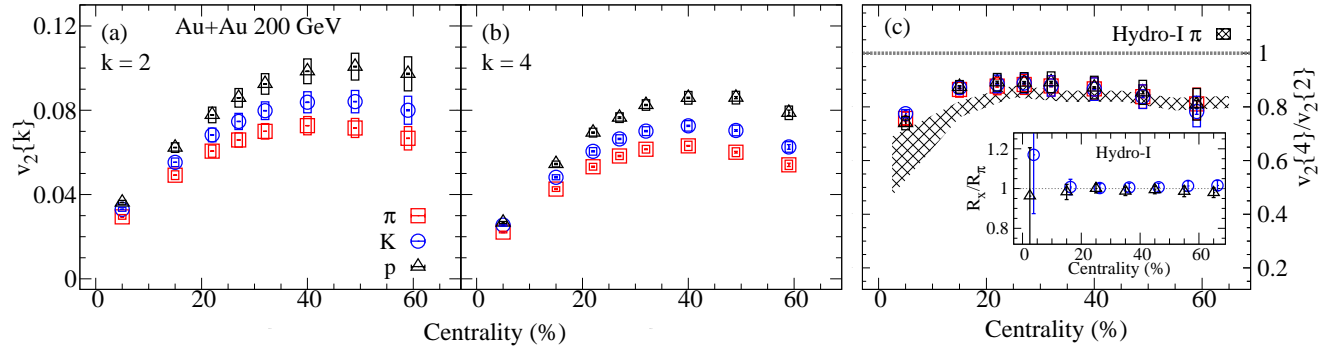}
 \vskip -0.4cm
\caption{Comparison of the centrality dependence of ${v_{2}\lbrace 2 \rbrace}$ (a), 
${v_{2}\lbrace 4 \rbrace}$ (b) and the ratio ${v_{2}\lbrace 4 \rbrace / v_{2}\lbrace 2 \rbrace}$ (c), 
for different particle species in the $p_{\rm T}$ range $0.2 - 2.0$~\GeVc for Au+Au collisions at \roots = 200~GeV. The vertical lines and the open boxes indicate the respective statistical and systematic uncertainties.
The hashed band in (c) shows the results for charged pions from Ref~\cite{Schenke:2019ruo} (Hydro-I). 
{\color{black}The inset shows the ratio $R_{K,p}/R_{\pi}$  $(R ={v_{2}\lbrace 4 \rbrace / v_{2}\lbrace 2 \rbrace})$ for the respective particle species.
 }}
\label{Fig:2}
\vskip -0.5cm
\end{figure*}
The non-flow contributions to the two-particle cumulants, that typically involve particles emitted within 
a localized region in ${\eta}$, can be mitigated via the two-subevents method~\cite{Zhou:2015iba,Jia:2017hbm,Magdy:2020bhd}. 
The associated two-particle cumulants can be expressed as: 
\begin{eqnarray}
 \langle\langle 2 \rangle\rangle_{a|b} &=& \langle\langle e^{in(\phi_{1}^a-\phi_{2}^b)} \rangle\rangle, \\
 c_n\{2\} &=& \left\langle\left\langle 2\right\rangle\right\rangle_{a|b},
 \label{eq:c3}
\end{eqnarray}
where Eqs.~(\ref{eq:c2-1})-(\ref{eq:c3}) lead to the following cumulant-based definitions for the two-, four-, and 
six-particle harmonic flow coefficients $v_n$:
\begin{eqnarray}
v_n\{2\} &=& \sqrt{c_n\{2\}},      \label{eq:c4-1} \\
v_n\{4\} &=&\sqrt[4]{-c_n\{4\}},   \label{eq:c4-2} \\
v_n\{6\} &=& \sqrt[6]{c_n\{6\}/4}. \label{eq:c4-3}
\end{eqnarray}
The subevents method was used to evaluate the two-particle cumulants for the non-overlapping ${\eta}$ interval  ${|\Delta\eta| > 0.6}$ ({\it i.e.} ${\eta^{a}} > 0.3$ for sub-event ${a}$ and ${\eta^{b}} < -0.3$ for sub-event ${b}$),  but not the four- and six-particle cumulants due to the limited acceptance and statistics of the measurements. Instead, `traditional’ four- and six-particle cumulants were obtained via the method with particle weights that reflect the efficiency and acceptance correction~\cite{Bilandzic:2010jr,Bilandzic:2013kga}.

For a Gaussian distribution of  the flow fluctuations, the fluctuations contributions to the {$\rm n^{th}$}-order flow harmonics can be written as~\cite{Voloshin:2008dg,Bhalerao:2011yg}:
\begin{eqnarray}
v_n\{2\} &\approx & \langle v_n \rangle + \sigma_n^{2}/(2\langle v_n \rangle), \label{eq:c5-1}\\
v_n\{4\} &\approx & \langle v_n \rangle -  \sigma_n^{2}/(2\langle v_n \rangle) ,\label{eq:c5-2}\\
v_n\{6\} &\approx & \langle v_n \rangle -  \sigma_n^{2}/(2\langle v_n \rangle). \label{eq:c5-3}
\end{eqnarray}
Eqs.~\ref{eq:c5-1}, \ref{eq:c5-2} and \ref{eq:c5-3} are also valid for other distributions in the limit that the variance $\sigma_n << \langle v_n \rangle$. 
In this work, the ratio between the four-particle elliptic flow $v_2\{4\}$, and  the two-particle non-flow-suppressed elliptic flow, $v_2\{2\}$ at a given centrality, is used to estimate the strength of the elliptic flow fluctuations' relative to the measured elliptic flow strength~\cite{Giacalone:2017uqx,Alba:2017hhe}. Note that $v_{2}\{4\}/ v_{2}\{2\} \approx 1.0$ indicates minimal, if any, fluctuations whereas $v_{2}\{4\}/ v_{2}\{2\} < 1.0$ indicates more significant fluctuations as this ratio decreases.

The presented measurements' systematic uncertainties are obtained from variations 
in the analysis cuts for event selection, track selection and non-flow suppression;
(i) event selection was varied via cuts on the vertex positions determined in the TPC along the beam 
direction,  $v_{z}$,   to  $v_{z}$ $>$ 0 cm and $v_{z}$ $<$ 0 cm. (ii) Track selection was varied by (a) reducing the {DCA from its nominal value of 3~cm to 2~cm,} and (b) increasing the number of TPC space points used from more than $15$ points to more than $20$ points. (iii) The pseudorapidity gap, $\Delta\eta~=~\eta_{1}-\eta_{2}$ for the track pairs, used to mitigate the non-flow effects  due to resonance decays, Bose-Einstein correlations, and the fragments of individual jets, was varied from $|\Delta\eta| > 0.6$ to $|\Delta\eta| > 0.8$. 

The $\Delta\eta$ cut does not entirely suppress possible long-range non-flow contributions (e.g., jets in a dijet event), which increase from central to peripheral events and decrease with beam energy. Estimates of the systematic uncertainty due to this  residual non-flow contribution can be made via several techniques~\cite{ATLAS:2012cix,STAR:2004jwm,Lacey:2020ime,STAR:2014qsy}. The peripheral subtraction method~\cite{ATLAS:2012cix}, which assumes that the long-range non-flow is independent of centrality, indicates uncertainties that range from 1\% in central collisions to 13\% in peripheral collisions at \roots= 200 GeV, and are included in the overall uncertainties.
Due to the lower jet yields for beam energies $\alt 63$~GeV \cite{STAR:2017ieb}, the much smaller associated uncertainties are not included in their respective overall systematic uncertainty estimate.

For identified particle species, the particle identification cuts were also varied about their nominal values~\cite{Adamczyk:2013gw}.
The overall systematic uncertainty for identified and inclusive charged hadrons, assuming independent sources, was estimated via a quadrature sum of the uncertainties resulting from the respective cut variations.  They range from 4\% to 15\% for $v_2\{2\}$,  2\% to 4\% for $v_2\{4\}$ and $v_2\{6\}$, and 4\% to 13\% for ${v_{2}\lbrace 4 \rbrace / v_{2}\lbrace 2 \rbrace}$, from central to peripheral collisions, depending on the beam energy. The non-flow-associated uncertainty dominates the overall uncertainty of ${v_{2}\lbrace 4 \rbrace / v_{2}\lbrace 2 \rbrace}$ since the effects of the other cut variations approximately cancel.

In Fig. \ref{Fig:1} the $p_{\rm T}$-integrated  two-, four-, and six-particle elliptic flow (a) and the ratio 
${v_{2}\lbrace 4 \rbrace / v_{2}\lbrace 2 \rbrace}$ (b), are presented as a function of centrality for Au+Au collisions at \roots = 200 GeV. 
 Note that  the range 1-5\% is used instead of 0-5\% \cite{STAR:2011ert} to ensure positive values for $v_2\{4\}$ in central collisions. Further study~\cite{Aaboud:2019sma} is required to understand this sign change fully.
Figure~\ref{Fig:1} (a) shows the known characteristic centrality dependence of two-, four- and six-particle elliptic flow, 
as well as quantitative agreement between $v_2\{4\}$ and $v_2\{6\}$. The difference between the magnitudes for 
$v_2\{2\}$ and those for $v_2\{4\}$ and $v_2\{6\}$ reflects the important role of the flow fluctuations.
The similarity between $v_2\{4\}$ and $v_2\{6\}$, within statistical uncertainties,  is consistent with a Gaussian hypothesis of the flow fluctuations.
%
%
The ratio $v_2\{4\}/v_2\{2\}$, presented in Fig.~\ref{Fig:1} (b), serves as a metric for elliptic flow fluctuations; it shows the expected decrease in the magnitude of the fluctuations from central to mid-central collisions,
reminiscent of the pattern observed for the initial-state eccentricity fluctuations, ${\varepsilon_{2}\lbrace 4 \rbrace / \varepsilon_{2}\lbrace 2 \rbrace}$ \cite{Alba:2017hhe}, shown by the blue dashed curve. 
 The hashed bands in Fig.~\ref{Fig:1} represent the results from two hydrodynamical model calculations~\cite{Alba:2017hhe,Schenke:2020mbo}. Hydro-I~\cite{Schenke:2010nt,Schenke:2019ruo} uses an IP-Glasma~\cite{Schenke:2012wb} inspired initial-state in conjunction with  the UrQMD~\cite{Bass:1998ca,Bleicher:1999xi} afterburner. It also imposes the effects of global momentum conservation and the local charge conservation. Hydro-II~\cite{Alba:2017hhe} employs the {T$\mathrel{\protect\raisebox{-2.1pt}{R}}$ENTo} model~\cite{Moreland:2014oya} initial-state and does not include the UrQMD afterburner. 
 Both models show good qualitative agreement with the $v_2$ data (Fig.~\ref{Fig:1} (a)). The data-model comparisons in Fig.~\ref{Fig:1} (b) indicate that  Hydro-II~\cite{Alba:2017hhe} over-predicts the measurements in mid-central and peripheral collisions, but Hydro-I~\cite{Schenke:2019ruo} is in good overall agreement with the presented measurements.
%
The hydrodynamic model predictions contrast with the corresponding eccentricity fluctuations (dashed blue line) which appear to under-predict the measured fluctuations in peripheral events.  The latter is to be expected if eccentricity fluctuations are not the only source of the flow fluctuations. However, possible residual non-flow contribution to $v_{2}\lbrace 2 \rbrace$ could also contribute to this difference. Nonetheless, the similarity between the centrality dependence of the ratio ${\varepsilon_{2}\lbrace 4 \rbrace / \varepsilon_{2}\lbrace 2 \rbrace}$, and that for the ${v_{2}\lbrace 4 \rbrace / v_{2}\lbrace 2 \rbrace}$ measurements, suggests that eccentricity fluctuations dominate the flow fluctuations in central and mid-central collisions.
%
%
 \begin{figure*}[!ht]
  \vskip -0.4cm
\includegraphics[width=1.0\linewidth,angle=0]{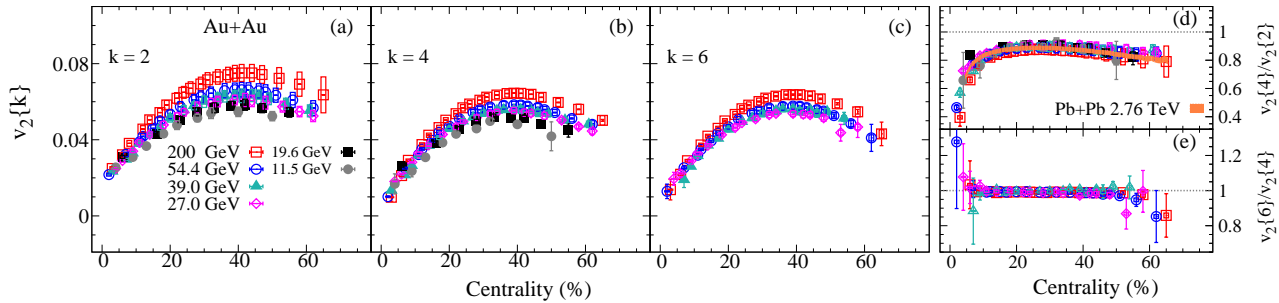}
 \vskip -0.4cm
\caption{Comparison of the centrality dependence of the charged hadrons  $v_2\{2\}$ (a), $v_2\{4\}$ (b), $v_2\{6\}$ (c) and the ratios ${v_{2}\lbrace 4 \rbrace / v_{2}\lbrace 2 \rbrace}$ (d) and ${v_{2}\lbrace 6 \rbrace / v_{2}\lbrace 4 \rbrace}$ (e), in the $p_{\rm T}$ range $0.2 - 4.0$~\GeVc for  Au+Au collisions at $\roots$ = 11.5 -- 200~GeV.
The vertical lines and the open boxes indicate the respective statistical and systematic uncertainties.
The shaded band in (d) indicates the ratios obtained from the LHC  measurements for the $p_{\rm T}$ range $0.2 -  3.0$~\GeVc for Pb+Pb collisions at \roots = 2.76~TeV~\cite{Acharya:2018lmh}; only statistical uncertainties are shown for the latter.}
\label{Fig:3}
\end{figure*}

%
%
 \begin{figure*}[t]
  \vskip -0.5cm
 \includegraphics[width=1.0\linewidth, angle=0]{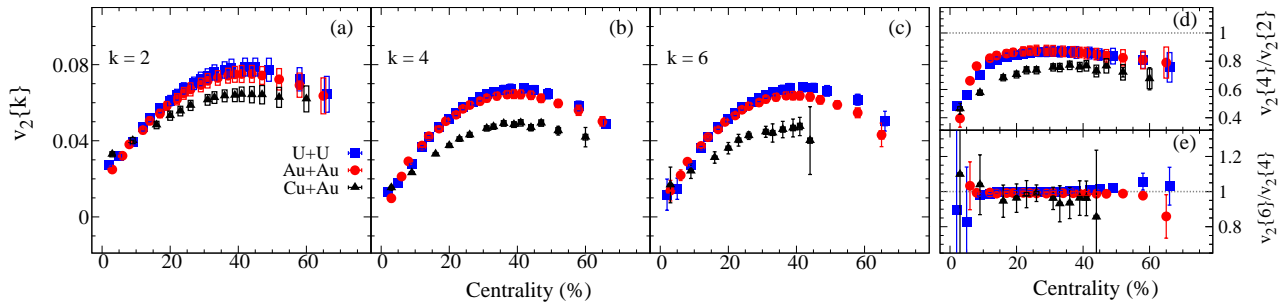}
 \vskip -0.4cm
 \caption{ Comparison of the centrality dependence of the charged hadrons  $v_2\{2\}$ (a), $v_2\{4\}$ (b), $v_2\{6\}$ (c) and the ratios ${v_{2}\lbrace 4 \rbrace / v_{2}\lbrace 2 \rbrace}$ (d) and ${v_{2}\lbrace 6 \rbrace / v_{2}\lbrace 4 \rbrace}$ (e), in the $p_{\rm T}$ range $0.2 - 4.0$~\GeVc for  U+U ($\roots$ = 193~GeV), Au+Au and Cu+Au collisions at $\roots$ = 200~GeV. The vertical lines and the open boxes indicate the respective statistical and systematic uncertainties.
 }
\label{Fig:4}
\vskip -0.5cm
 \end{figure*}
The species-dependence of the flow fluctuations can give insight on possible contributions from other fluctuation sources~\cite{Martinez:2019jbu,Magdy:2020gxf,CMS:2021qqk}. Here, an essential point is that the fluctuations generated during the hadronization of the QGP could lead to a difference in the magnitude of the fluctuations for different particle species. Figure~\ref{Fig:2} shows a comparison of the measured centrality dependence of $v_2\{2\}$ (a), $v_2\{4\}$ (b) and the ratio $v_2\{4\}/v_2\{2\}$ (c), for pions,  kaons, and protons in Au+Au collisions at $\roots$ = 200~GeV; 
for these $v_2\{4\}$ results, a procedure which uses one identified hadron and three inclusive charged hadrons was employed.
The $v_2\{2\}$  and $v_2\{4\}$ measurements exhibit the well known mass ordering~\cite{Hirano:2007ei} for these particle species.

Figure~\ref{Fig:2}~(c) compares  the ${v_{2}\lbrace 4 \rbrace / v_{2}\lbrace 2 \rbrace}$ ratios for pions, kaons and protons;
they indicate that the magnitude and trend of the flow-fluctuations are independent of the particle-species.
The effects of mass ordering, apparent in Figs.~\ref{Fig:2} (a) and (b), are expected to cancel in these ratios~\cite{Alba:2017hhe,Magdy:2020gxf}, but the fluctuations might  not. 
Strikingly similar species independent patterns can also be seen for  the ratios obtained from the Hydro-I 
calculations~\cite{Schenke:2020mbo}, shown by the hatched band and the inset in Fig.~\ref{Fig:2} (c). A similar species independent result is obtained for  Hydro-II~\cite{Alba:2017hhe}, albeit with different magnitudes for the ${v_{2}\lbrace 4 \rbrace / v_{2}\lbrace 2 \rbrace}$ ratios.
A species independent ${v_{2}\lbrace 4 \rbrace / v_{2}\lbrace 2 \rbrace}$ ratio is expected 
if initial-state fluctuations dominate over other sources of fluctuations. 

The beam-energy dependence of the flow fluctuations can give insight into possible fluctuation sources associated with the expansion dynamics. Consequently, the {\color{black}flow and flow-fluctuation measurements} were performed  for Au+Au collisions spanning the range $\roots = 11.5-200$~GeV.
Figure~\ref{Fig:3} gives a summary of the centrality dependence of $v_2\{2\}$ (a), $v_2\{4\}$ (b), $v_2\{6\}$ (c) and the ratios ${v_{2}\lbrace 4 \rbrace / v_{2}\lbrace 2 \rbrace}$ (d) and ${v_{2}\lbrace 6 \rbrace / v_{2}\lbrace 4 \rbrace}$ (e) for the respective beam energies indicated. Figs.~\ref{Fig:3} (a) - \ref{Fig:3} (c) show an increase with increasing beam energy for the values of $v_2\{2\}$, $v_2\{4\}$, and  $v_2\{6\}$, that reflects the change in the expansion dynamics. 

The ${v_{2}\lbrace 4 \rbrace / v_{2}\lbrace 2 \rbrace}$  ratios shown in Fig.~\ref{Fig:3} (d) suggest that within the given uncertainties, the flow fluctuations are weakly dependent on  the beam energy, if at all, irrespective of the collision centrality. The magnitude and trend of these ratios are also comparable to those for the LHC  measurements for Pb+Pb collisions at \roots = 2.76~TeV~\cite{Acharya:2018lmh} and to the ${\varepsilon_{2}\lbrace 4 \rbrace / \varepsilon_{2}\lbrace 2 \rbrace}$ ratios, in central to mid-central collisions, shown in Fig.~\ref{Fig:1} (b). They suggest that the flow fluctuations associated with the expansion dynamics do not 
change substantially over the beam energy range  $\roots = 11.5-2760$~GeV. The comparable magnitudes for  ${v_{2}\lbrace 4 \rbrace / v_{2}\lbrace 2 \rbrace}$ and ${\varepsilon_{2}\lbrace 4 \rbrace / \varepsilon_{2}\lbrace 2 \rbrace}$ also suggest that the initial-state eccentricity fluctuations dominate  the flow fluctuations encoded in the ratio ${v_{2}\lbrace 4 \rbrace / v_{2}\lbrace 2 \rbrace}$. The ${v_{2}\lbrace 6 \rbrace / v_{2}\lbrace 4 \rbrace}$  ratios in Fig.~\ref{Fig:3} (e)  indicate values which are systematically less than one, but with little if any, dependence on centrality.  The  latter pattern, which contrasts with similar LHC measurements~\cite{ATLAS:2014qxy,CMS:2017glf,ALICE:2018rtz},  could be a further indication for the Gaussian-like nature of the flow fluctuations across the presented beam energies.

Further knowledge on the fluctuation sources can be obtained by comparing  the measurements
for collisions of U+U, Au+Au and Cu+Au  at similar collision energy. Here, it is noteworthy that
the prolate deformation of uranium, the oblate deformation of Au, and the asymmetry and system size
for Cu+Au collisions, can lead to different initial-state eccentricities for the same centrality, especially in central collisions. Figure~\ref{Fig:4} shows a summary of the centrality dependence of $v_2\{2\}$ (a), $v_2\{4\}$ (b), $v_2\{6\}$ (c) and the ratios ${v_{2}\lbrace 4 \rbrace / v_{2}\lbrace 2 \rbrace}$ (d) and ${v_{2}\lbrace 6 \rbrace / v_{2}\lbrace 4 \rbrace}$~(e) for U+U ($\roots = 193$~GeV), Au+Au and Cu+Au collisions at $\roots = 200$~GeV. Figures~\ref{Fig:4} (a) - \ref{Fig:4} (c) provide a clear indication that $v_2\{2\}$, $v_2\{4\}$, and  $v_2\{6\}$ are system-dependent and follows a system-size hierarchy with more pronounced differences for Cu+Au.  This system-size hierarchy can be attributed to the system-dependent eccentricity hierarchy~\cite{Schenke:2014tga}.

The ${v_{2}\lbrace 4 \rbrace / v_{2}\lbrace 2 \rbrace}$ ratios shown in Fig.~\ref{Fig:4} (d), indicate the expected decrease in the magnitude of the fluctuations from central to peripheral collisions for all three systems. However, in contrast to the energy dependence shown in Fig.~\ref{Fig:3} (d), the system dependence of the flow fluctuations is now apparent, albeit with a much smaller difference between U+U and Au+Au than the difference between Cu+Au and U+U or Au+Au. These results point to an increasingly important role for flow fluctuations as the system size is reduced. The magnitude and trends of the ${v_{2}\lbrace 6 \rbrace / v_{2}\lbrace 4 \rbrace}$  ratios in Fig.~\ref{Fig:4} (e)  are similar to those in Fig.~\ref{Fig:3} (e), suggesting that the Gaussian-like nature of the flow fluctuations is system-independent.

In summary, we have used the two- and multiparticle cumulants method to make comprehensive measurements of two-, four-, and six-particle elliptic flow  and flow fluctuations in collisions of U+U at \roots = 193~GeV, Cu+Au at \roots = 200~GeV and Au+Au for the range \roots = 11.5 - 200~GeV. 
 The measurements show the expected characteristic dependence of ${v_{2}\lbrace 2 \rbrace}$, ${v_{2}\lbrace 4 \rbrace}$ and ${v_{2}\lbrace 6 \rbrace}$ on centrality, system size and beam energy. 
The elliptic-flow fluctuations extracted from these measurements indicate more  substantial fluctuations in more central collisions, a dependence on collision system, and  little if any dependence on particle species and beam energy.
Comparisons of these results to similar LHC measurements, as well as to viscous hydrodynamical calculations and T$\mathrel{\protect\raisebox{-2.1pt}{R}}$ENTo model eccentricity ratios, suggest that initial-state-driven fluctuations dominate the flow fluctuations in the collisions studied.
A complete set of model comparisons to this comprehensive data set is needed to flesh out the detailed initial- and final-state-driven contributions to flow fluctuations. The mapping of such contributions could serve to discern between different initial-state models, as well as constrain the fluctuations-related uncertainties associated with the extraction of ${\eta}/{s} (T, \mu_B)$.

\vskip 1cm
\begin{acknowledgements}
%
We thank the RHIC Operations Group and RCF at BNL, the NERSC Center at LBNL, and the Open Science Grid consortium for providing resources and support.  This work was supported in part by the Office of Nuclear Physics within the U.S. DOE Office of Science, the U.S. National Science Foundation, National Natural Science Foundation of China, Chinese Academy of Science, the Ministry of Science and Technology of China and the Chinese Ministry of Education, the Higher Education Sprout Project by Ministry of Education at NCKU, the National Research Foundation of Korea, Czech Science Foundation and Ministry of Education, Youth and Sports of the Czech Republic, Hungarian National Research, Development and Innovation Office, New National Excellency Programme of the Hungarian Ministry of Human Capacities, Department of Atomic Energy and Department of Science and Technology of the Government of India, the National Science Centre and WUT ID-UB of Poland, the Ministry of Science, Education and Sports of the Republic of Croatia, German Bundesministerium f\"ur Bildung, Wissenschaft, Forschung and Technologie (BMBF), Helmholtz Association, Ministry of Education, Culture, Sports, Science, and Technology (MEXT) and Japan Society for the Promotion of Science (JSPS).

\end{acknowledgements}
\bibliography{ref} 
\end{document}